\documentclass[twocolumn,tighten,times]{aastex62}
\usepackage{mathrsfs}
\usepackage{amsmath}
\usepackage{url}
\usepackage[normalem]{ulem}
\usepackage{graphicx}
\usepackage{float}
\turnoffedit

\usepackage{natbib}
\usepackage{color}
\bibpunct{(}{)}{;}{a}{}{,}

\newcommand\aproxgt{\mathrel{%
      \rlap{\raise 0.511ex \hbox{$>$}}{\lower 0.511ex \hbox{$\sim$}}}}
\newcommand\aproxlt{\mathrel{%
      \rlap{\raise 0.511ex \hbox{$<$}}{\lower 0.511ex \hbox{$\sim$}}}}

\newcommand{\ignore}[1]{}

\newcommand{\swift}{{\it Swift}}
\newcommand{\hst}{{\it HST\,}}

\newcommand{\et}{{et~al.\, }}

\newcommand{\civ}{\mbox{\rm C\,{\sc iv}}}

\newcommand{\siIVoIV}{\mbox{\rm Si\,{\sc iv}$+$O\,{\sc iv}]}}
\newcommand{\HeIIoIII}{\mbox{\rm He\,{\sc ii}$+$O\,{\sc iii}]}}

\shorttitle{A wind-based unification model for NGC5548}
\shortauthors{Dehghanian \et}

\begin{document}

\title{A wind-based unification model for NGC 5548:\linebreak 
spectral holidays, non-disk emission, and implications for changing-look quasars}

\author{M.~Dehghanian}
\affiliation{\ignore{UKy}Department of Physics and Astronomy, The University of Kentucky, Lexington, KY 40506, USA}

\author{G.~J.~Ferland}
\affiliation{\ignore{UKy}Department of Physics and Astronomy, The University of Kentucky, Lexington, KY 40506, USA}

\author{B.~M.~Peterson}
\affiliation{Space Telescope Science Institute, 3700 San Martin Drive, Baltimore, MD 21218, USA}
\affiliation{Department of Astronomy, The Ohio State University, 140 W 18th Ave, Columbus, OH 43210, USA}
\affiliation{Center for Cosmology and AstroParticle Physics, The Ohio State University, 191 West Woodruff Ave, Columbus, OH 43210, USA}

\author{G.~A.~Kriss}
\affiliation{Space Telescope Science Institute, 3700 San Martin Drive, Baltimore, MD 21218, USA}

\author{K.~T.~Korista}
\affiliation{\ignore{WM}Department of Physics, Western Michigan University, 1120 Everett Tower, Kalamazoo, MI 49008-5252, USA}

\author{M.~Chatzikos}
\affiliation{\ignore{UKy}Department of Physics and Astronomy, The University of Kentucky, Lexington, KY 40506, USA}

\author{F.~Guzm\'{a}n}
\affiliation{\ignore{UKy}Department of Physics and Astronomy, The University of Kentucky, Lexington, KY 40506, USA}

\author{N.~Arav}
\affiliation{Department of Physics, Virginia Tech, Blacksburg, VA 24061, USA}

\author{G.~De~Rosa}
\affiliation{Space Telescope Science Institute, 3700 San Martin Drive, Baltimore, MD 21218, USA}

\author{M.~R.~Goad}
\affiliation{\ignore{Leicester}Department of Physics and Astronomy, University of Leicester,  University Road, Leicester, LE1 7RH, UK}

\author{M.~Mehdipour}
\affiliation{\ignore{SRON}SRON Netherlands Institute for Space Research, Sorbonnelaan 2, 3584, CA Utrecht, The Netherlands}

\author{P.~A.~M.~van~Hoof}
\affiliation{\ignore{Belgium}Royal Observatory of Belgium, Ringlaan 3, B-1180 Brussels, Belgium}


\begin{abstract}
The 180-day Space Telescope and Optical Reverberation Mapping campaign on NGC 5548 discovered an anomalous period, the broad-line region (BLR) holiday, in which the emission lines decorrelated from the continuum variations. This is important since the correlation between the continuum-flux variations and the emission-line response is the basic assumption for black hole (BH) mass determinations through reverberation mapping. During the BLR holiday the high-ionization intrinsic absorption lines also decorrelated from the continuum  as a result of variable covering factor of the line of sight (LOS) obscurer. The emission lines are not confined to the LOS, so this does not explain the BLR holiday. If the LOS obscurer is a disk wind, its streamlines must extend down to the plane of the disk and the base of the wind would lie between the BH and the BLR, forming an equatorial obscurer. This obscurer can be transparent to ionizing radiation, or can be translucent, blocking only parts of the SED, depending on its density. An emission-line holiday is produced if the wind density increases only slightly above its transparent state. Both obscurers are parts of the same wind, so they can have associated behavior in a way that explains both holidays. A very dense wind would block nearly all ionizing radiation, producing a Seyfert 2 and possibly providing a contributor to the changing-look AGN phenomenon. Disk winds are very common and we propose that the equatorial obscurers are too, but mostly in a transparent state.
\end{abstract}


2\section{INTRODUCTION } 

AGN STORM, the AGN Space Telescope and Optical Reverberation Mapping project, is the 
largest spectroscopic reverberation mapping (RM) campaign to date. NGC 5548 
\edit1{was observed }with \hst/\emph{COS} \edit1{nearly}daily over six months in 2014 \citep{DeRosa15, Edelson15, Fausnaugh16, Goad16, Pei17, Starkey17, Mathur17}, with the goal of determining the kinematics and geometry of the central regions using RM methods. 
\cite{Goad16}, hereafter G16, revealed some unexpected results: 
about 60 days into the observing campaign, the FUV continuum and broad emission line variations, 
which are typically highly correlated and form the basis of RM, became “decorrelated” 
for $\sim$ 60-70 days, after which time the emission lines returned to their normal behavior. During this time, the equivalent widths (EWs) of the emission lines dropped by at most 25-30\%.
This anomalous behavior, hereafter the ``emission-line holiday'', 
was investigated by G16, \cite{Pei17,Mathur17,sun18} among others, although 
no physical model \edit1{to explain it} has been proposed. The occurrence of the emission-line holiday shows 
that we are missing an important part of the physics of the inner regions of AGN.

As discussed by \cite{Kriss19} and \citet[hereafter D19]{Deh19} the same holiday happened approximately simultaneously (within measurement uncertainties) for the high-ionization narrow intrinsic absorption lines. D19 shows that changes in the covering factor (CF) of the line of sight (LOS) obscurer \citep{Kaastra14} explains the absorption-line holiday. The SED emitted by the source passes through this obscurer and then ionizes the absorbing clouds. Depending on the LOS CF of the obscurer, the transmitted SED changes in a way that reproduces the decorrelated behavior in some absorption lines.  The LOS CF deduced from \swift \ observations confirms this \edit1{hypothesis} (D19). 

Here, we examine the physics by which a related emission-line holiday could occur.  We take the obscurer to be a wind launched from the accretion disk, with variable mass-loss rate and hydrogen density. Figure 1 shows a cartoon with one possible geometry.  We show that for low hydrogen densities the obscurer near the disk is almost transparent and so has no effect on the SED striking the BLR. However, for higher densities it can obscure much of the ionizing radiation, producing the emission-line holiday. In this case, the observed UV continuum is not a good proxy for the ionizing flux.  Finally, for even higher gas densities, little ionizing radiation strikes the BLR. In this case, broad-line emission is strongly suppressed, resulting in something like a changing-look AGN. We suggest that an equatorial obscurer associated with a disk wind produces the BLR holiday, and may in more extreme circumstances contribute to causing a changing-look AGN.  

In Section 2, we set up a simple model of the BLR with no obscurer. Section 3 investigates how changes in the equatorial obscurer's hydrogen density change the transmitted SED. We then show, in Section 4, that the BLR responds to this variable equatorial obscurer in agreement with observations.  
Small changes in the obscurer's density reproduce the emission-line holiday and account for the amplitude of the variability in various lines. 
\edit2{If the covering fraction of the LOS obscurer also increases as the 
equatorial obscurer becomes more substantial, 
a simultaneous absorption-line holiday will be produced.}
  
\section{A baseline BLR with changing luminosity  }

\begin{figure}[H]
\centering
\begin{minipage}{3 in}
\includegraphics [width=3 in]{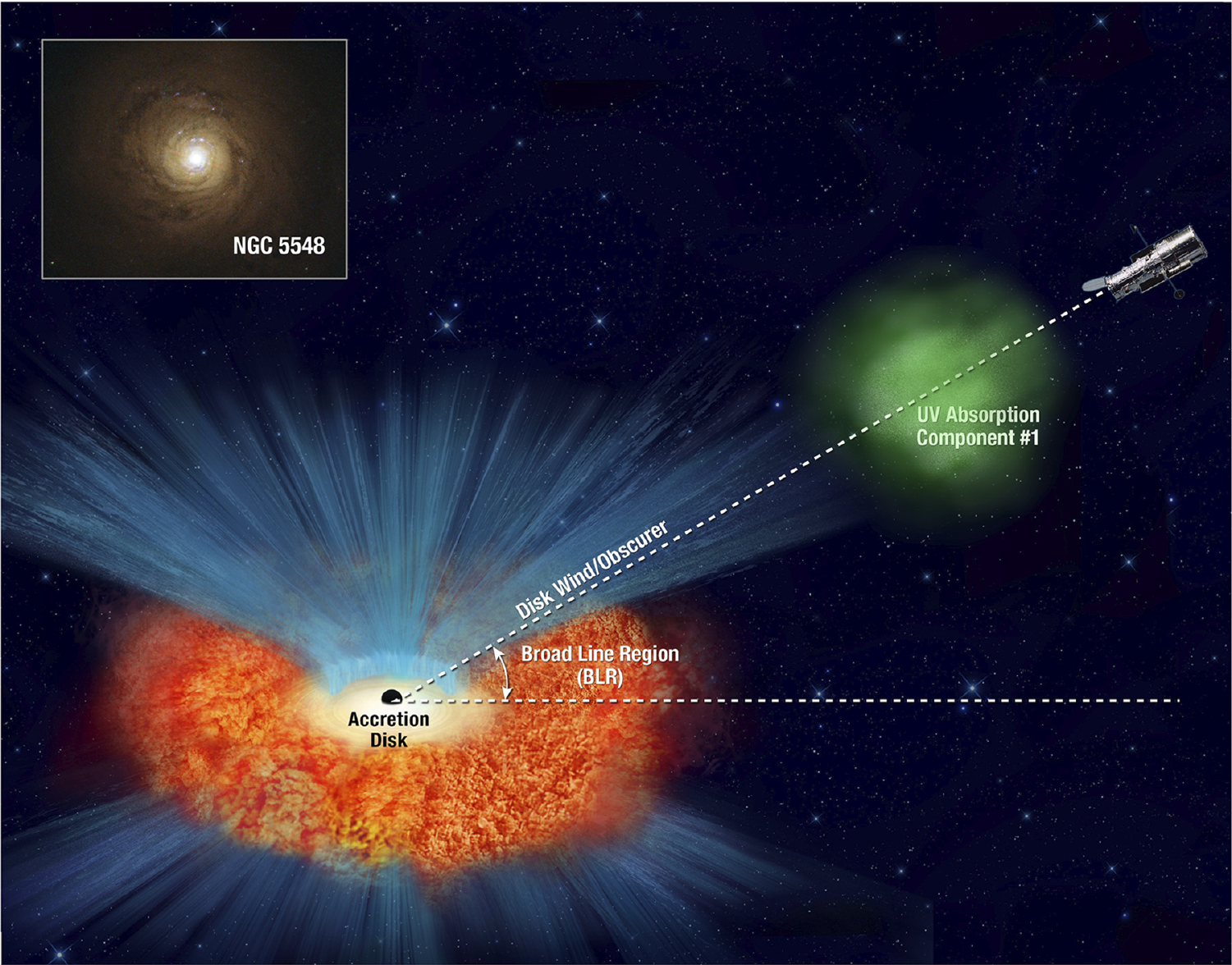}
\end{minipage}
 \caption{Diagram of the disk wind in NGC 5548 (not to scale). The BH is surrounded by the accretion disk. At larger radii the BLR is indicated by orange/red turbulent clouds. The disk wind rises nearly vertically from the surface of the accretion disk, where it has a dense, high-column-density base. At higher elevations, radiation pressure accelerates the wind and bends the streamlines down along the ~30 degree inclination of the observer’s LOS \edit2 {to the rotation axis of the disk \citep{Kaastra14}}.}\label{f1}
\end{figure}

Figure~\ref{f1} shows the geometry of the central regions, including the obscurer, based on \cite{Kaastra14} figure 4. We note that the \cite{Kaastra14} figure only highlights the portion of the disk wind that forms the obscurer along our LOS.
The critical differences in our illustration in Figure~\ref{f1} are (1)~we show the disk wind as an axisymmetric structure, (2) we show the full wind, with streamlines tracing from the surface of the disk to the gas lying along our LOS, and (3) we locate the obscurer interior to the BLR. The LOS obscurer is the upper part of the wind, and we refer to the lower part as the ``equatorial obscurer''. 

Although \edit2{some of} the properties of the LOS obscurer are known \edit2{(such as its column density and x-ray absorption)}, there is no way to \edit1{determine} the properties of the obscurer near the disk. The density at the base is likely to be higher than at higher altitudes and the column density through the base of the wind toward the BLR is higher than along the LOS, and \edit2{therefore the wind is} potentially opaque. Although our LOS samples only a specific sight  line through the wind, \edit2{we assume} the structure along all other sight lines is comparable and therefore can affect the whole of the BLR. The obscurer has persisted over at least four years \citep{Mehd16}. If it is located interior to the BLR at $<0.5$ light days, where the orbital timescale is only 40 days, this longevity implies that the wind extends a full 360-degrees around the black hole. It thus forms an axisymmetric, cylindrical continuous flow around the BH and so always fully shields the BLR. For this reason, it is not likely that a changing CF of the equatorial obscurer could explain the broad emission-line holiday as well.

Here we develop a baseline model for the BLR to investigate how its  emission lines are affected by the variations of the SED striking it. At this stage, we avoid including the equatorial obscurer in our modeling, so changes in the emission-line spectrum are caused by the variations of the luminosity of the source. For simplicity, we do not model a full LOC\footnote{Locally optimally emitting clouds} similar to figure 2 of \cite{Korista00}.  Our baseline model is sufficient for the goal of this paper, which is to test how changes in the equatorial obscurer change the observed EW of the broad emission lines. We use the development version of Cloudy (C17), last described by \cite{Ferland17}, for all the photoionization models presented here.

To model the BLR, we fix its hydrogen column density to be $N$(H)=10$^{23}$ cm$^{-2}$, choose a hydrogen density of $n$(H)=10$^{11}$ cm$^{-3}$, and use solar abundances \citep{Ferland17}.  These are all typical values for the BLR \citep[following][]{Ferland92, Goad98,Kaspi99}. The remaining parameter is the flux of hydrogen ionizing photons $\phi$(H) (ionizing photons cm$^{-2}$ s$^{-1}$) striking the cloud.  For a given SED shape \citep[we use that of][as discussed by D19]{Mehd15} and location of the BLR, this flux depends on the luminosity, so changes in the flux simulate changes in the luminosity.
\edit2{We assume thermal line broadening evaluated for the gas kinetic 
temperature and atomic weight of each species.}

\begin{figure*}
\centering
\includegraphics [width=\textwidth]{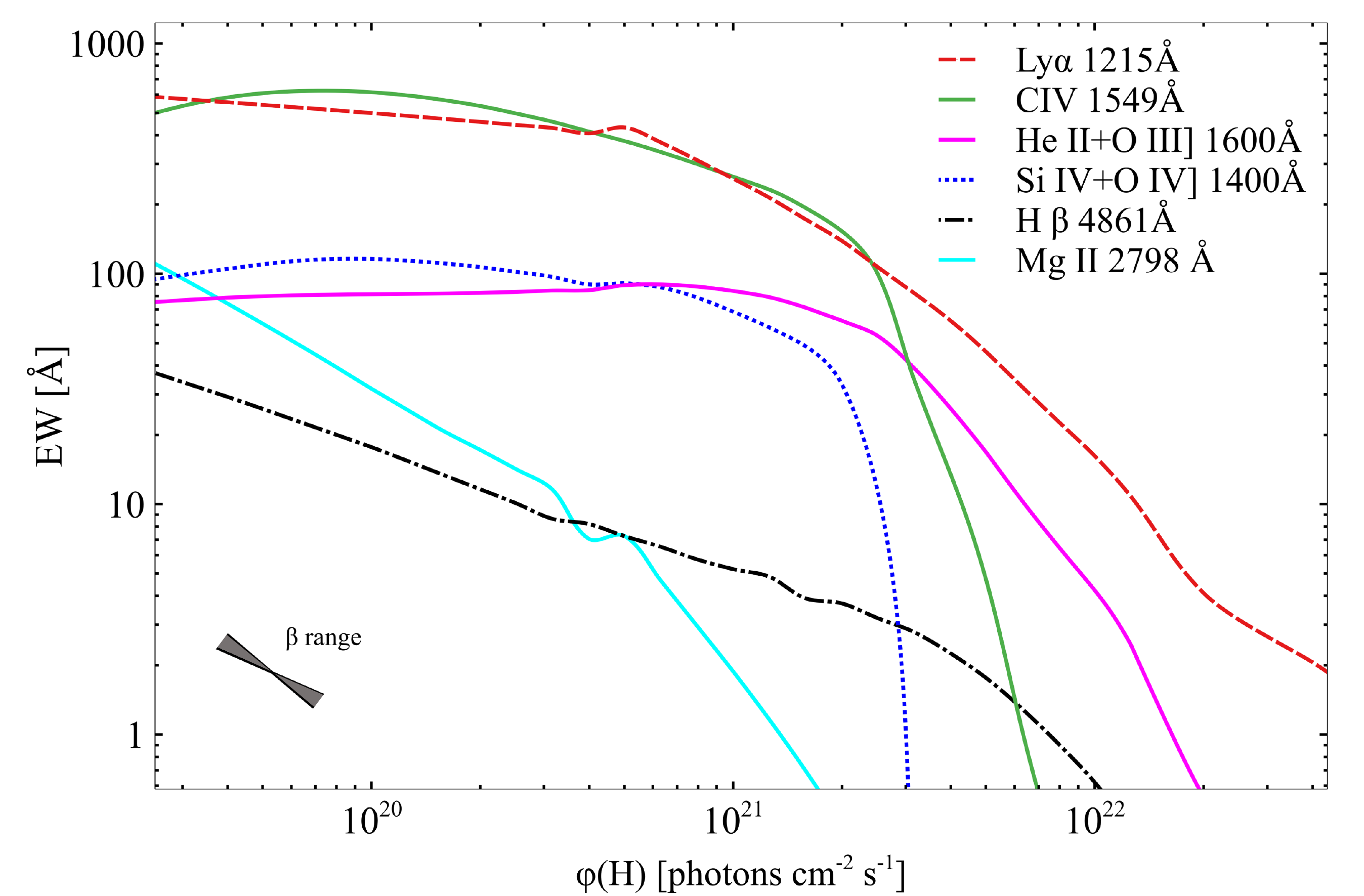}
\caption{EW of emission lines vs. the flux of hydrogen ionizing photons. The EWs are normalized to the continuum at 1367\AA.  \edit1{\edit2{For most of the lines,} the predicted EWs decrease when $\phi$(H)$ > 10^{20}$, the observed behavior. The bow tie shows the range of $\beta$ observed for various lines before the holiday.}}\label{f2}
\end{figure*}

\edit1{
The line EWs were observed to decrease as the luminosity increased before the holiday.
Figure~\ref{f2} shows our predicted EWs. 
The observations report a slope $\beta$ that fits  EW $\propto$ $L^{\beta}$.
G16 find  $\beta$ in the range -0.48 to -0.75 for Ly$\alpha$, Si IV+O IV],  \civ, and He~II+O~III],
while \cite{Pei17} find $\beta= -0.85$ for H$\beta$.  
This range of $\beta$ values is shown as the bow tie in the lower left corner.
Each of these lines has its own reverberation timescale, formation radius, and value of $\phi$(H).
Future work will examine using EW and $\beta$ to better constrain LOC models.
}

As Figure~\ref{f2} shows, variations of the luminosity can dramatically affect the BLR.   For $\phi$(H)$>10^{20}$ cm$^{-2}$ s$^{-1}$ the \civ \  EW, shown in green, behaves as in G16's figure 1b. Changes in the EW of \civ \ and H$\beta$ are consistent with G16 and \cite{Pei17}. 

In the next Section, we consider the effects of the equatorial obscurer on the BLR. 
To do this, we only change the parameters of the obscurer, while we freeze all  BLR parameters, including the unobscured flux,
which we take to be $\phi($H$)=10^{20}$ cm$^{-2}$ s$^{-1}$. Our goal is only to demonstrate a scenario that produces emission-line holidays, so we are not trying to fine tune the parameters.

\section{The SED transmitted through the equatorial obscurer}

As Figure~\ref{f1} shows, we assume that the obscurer is a wind extending from the equator to at least our LOS. This means that the BLR is ionized by the SED transmitted through the lowest part of the wind, the equatorial obscurer.  Here we investigate how the SED transmitted through the equatorial obscurer changes as the wind parameters change. 

There are no observational constraints on the equatorial obscurer, but it seems likely that it is denser, perhaps with a larger column density, than \edit2{the more distant LOS obscurer.}. For simplicity, we hold its column density fixed at $N$(H)$=10^{23}$ cm$^{-2}$ and assume solar abundances. Since the broad UV absorption associated with the LOS obscurer partially covers the BLR and has velocities ($\sim 1500~\rm km~s^{-1}$) typical of the BLR \citep{Kaastra14}, we assume that the LOS obscurer is near or coincident with the outer portion of the BLR.  The equatorial obscurer must be closer to the black hole since it is launched from the disk. We choose $\phi($H$)=10^{20.3}$ cm$^{-2}$ s$^{-1}$, twice that of the BLR,
placing the obscurer at $r_{obscurer}=0.7\times r_{BLR}$. 
We do not know the exact location of the equatorial obscurer and these values are  chosen based \edit1{only} on the fact that it must be inside the BLR.

\edit1{As in D19, we are trying to identify the phenomenology that makes the 
observed changes possible and not to model any particular observation (section 3.3 of that paper).  
We wish to see how the changes in the optical depth of the intervening wind affects emission from the BLR.  
These changes could be caused by variations in the physical thickness of the wind, 
its density, the AGN luminosity, or the distance from the black hole. 
For simplicity we vary only one of these, 
the density, while keeping the others fixed.  As discussed in following sections, 
this change, while simple, does serve to illustrate the types of SEDs that will filter through the wind. }
 
\begin{figure*}
\centering
\includegraphics [width=\textwidth]{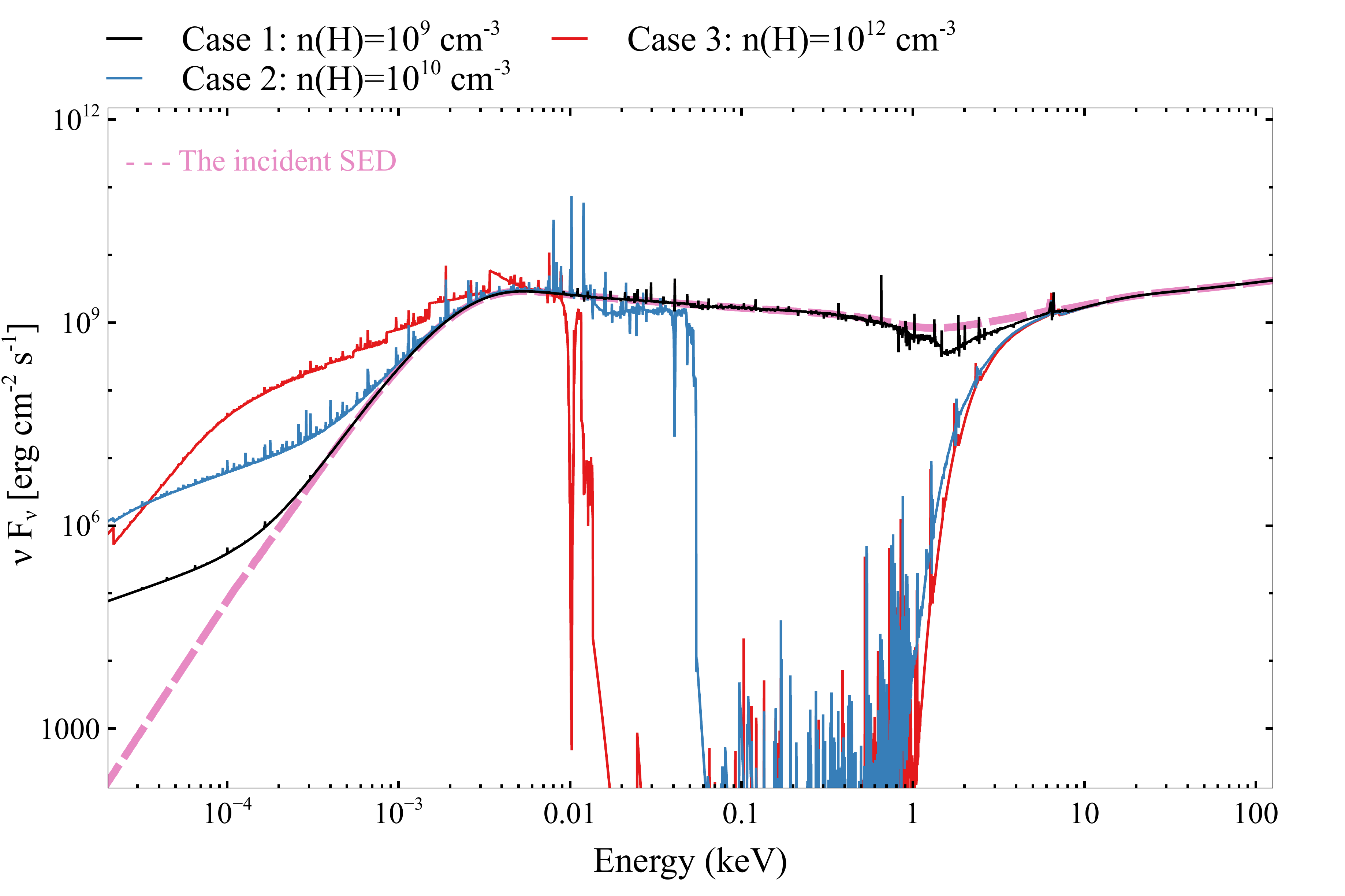}
 \caption{The SED transmitted through equatorial obscurer and incident upon the BLR is shown 
 for three different values of the hydrogen density. 
 \edit1{The unextinguished SED is also shown. The SED is dramatically dependent on the hydrogen density of the obscurer. High hydrogen densities produce strong \edit2{absorption in XUV region and strong emission in FUV/optical regions}. }}\label{f3}
\end{figure*}

 Changes in the mass-loss rate of the wind can cause changes in the hydrogen density of the equatorial obscurer.  We examine the effects of such variations upon the transmitted SED in Figure~\ref{f3}, which shows three typical SEDs. As the Figure shows, the shape of the SED is highly sensitive to the value of the hydrogen density.

The density and flux parameters chosen here do not matter in detail.  The transmitted SED actually depends on the ionization parameter, which is the ratio of the ionizing flux to the hydrogen density \citep{Osterbrock06}. Increasing the hydrogen density lowers the ionization parameter inversely.  
Particular values \edit2{of the density and flux}  do not matter as long as the ratio giving the ionization parameter is kept constant.

As the ionization parameter increases the level of ionization of the gas increases.  The gas opacity decreases as the number of bound electrons decreases. The ionization structure changes in ways that produce the three characteristic SEDs shown in Figure~\ref{f3}. These are the three cases: 

\begin{itemize}
\item Case1 has the lowest density and the highest ionization, and is shown in black.  This \edit1{wind} is fully ionized, has no H or He ionization fronts, and nearly fully transmits the entire incident SED. 
\item Case 2 has an intermediate density and is shown in blue. This has a He$^{2+}$ - He$^{+}$ ionization-front but no H ionization-front.  The incident SED is heavily absorbed for the \edit1{XUV energies\footnote{We refer to the region 6 -- 13.6 eV (912\,\AA\ to 2000\,\AA)
as FUV; 13.6 -- 54.4 eV (228\,\AA\ to 912\,\AA) as EUV; and 54.4 eV
to few hundred eV (less than 228\,\AA) as XUV.} }, although most of the hydrogen-ionizing radiation is transmitted. 
\item •	Finally, Case 3 is shown with the red line and has the highest density.  The \edit1{wind} has both  H and He ionization-fronts, and much of the light \edit 1{in the EUV and XUV regions} is absorbed.  
\end{itemize}

\section{The response of the BLR to changes of the transmitted continuum}
We now show how the EWs of the BLR lines in Figure~\ref{f2} are affected by changes in the transmitted SED of the equatorial obscurer. Figure~\ref{f4} shows how the EW of the strongest observed lines react as the density, $n$(H), of the equatorial obscurer varies. These \edit2{changes} are due to variations in the SED filtering through the equatorial obscurer. The three general types of SED shown in Figure~\ref{f3} produce the three different BLR regimes shown in Figure~\ref{f4}.  We examine each of these three cases in more detail:

\begin{figure*}
\centering
\includegraphics [width=\textwidth]{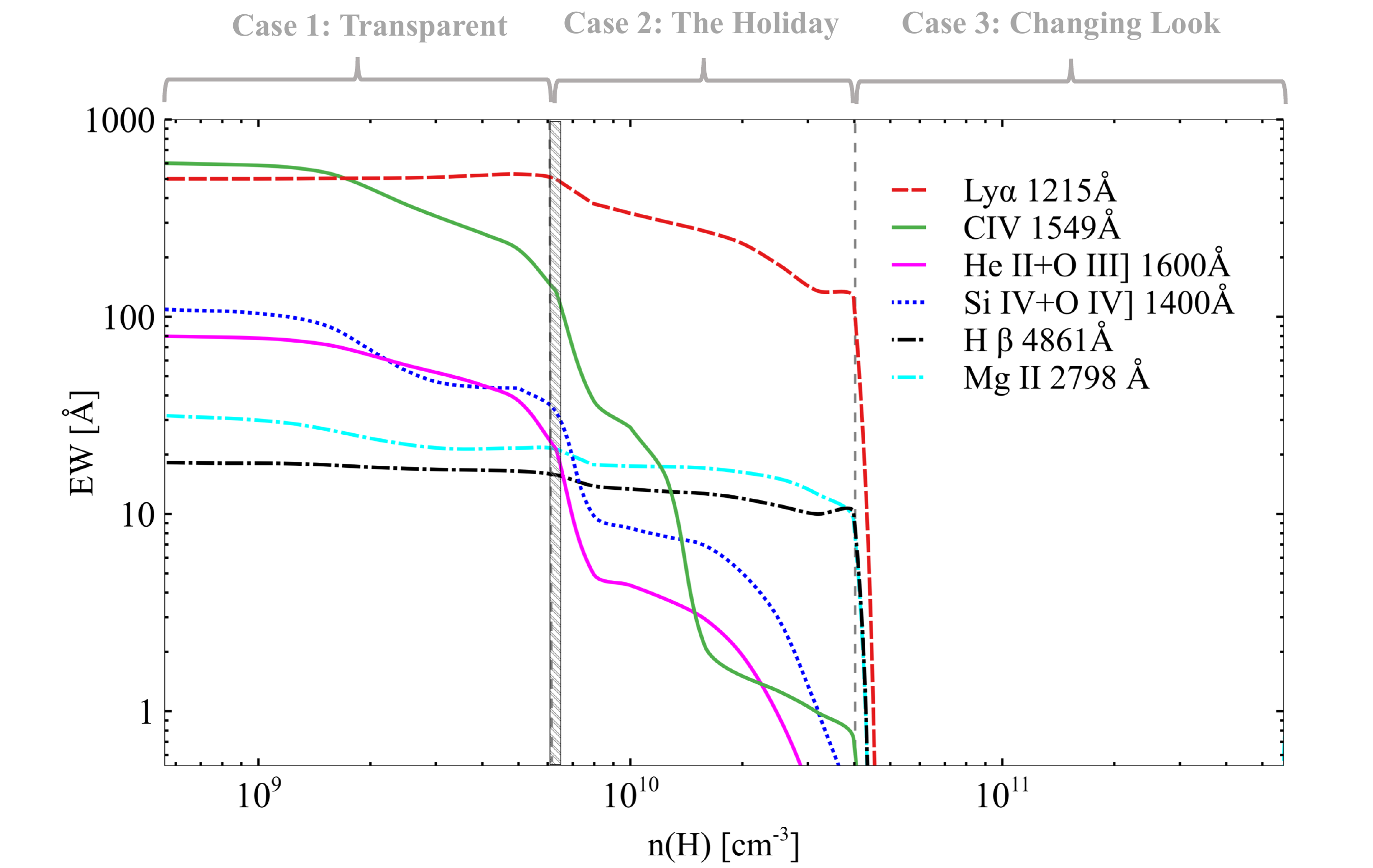}
 \caption{EW of all observed emission lines and \edit1{Mg II}  versus the density of the equatorial obscurer. \edit1{The plot is divided into three different cases, labeled at the top, based on the behavior of the EWs. 
 The cases are described in the text.  }The shaded area shows the range in which the holiday observed in NGC 5548 will be produced.}\label{f4}
\end{figure*}

\begin{figure*}
\centering
\includegraphics [width=\textwidth]{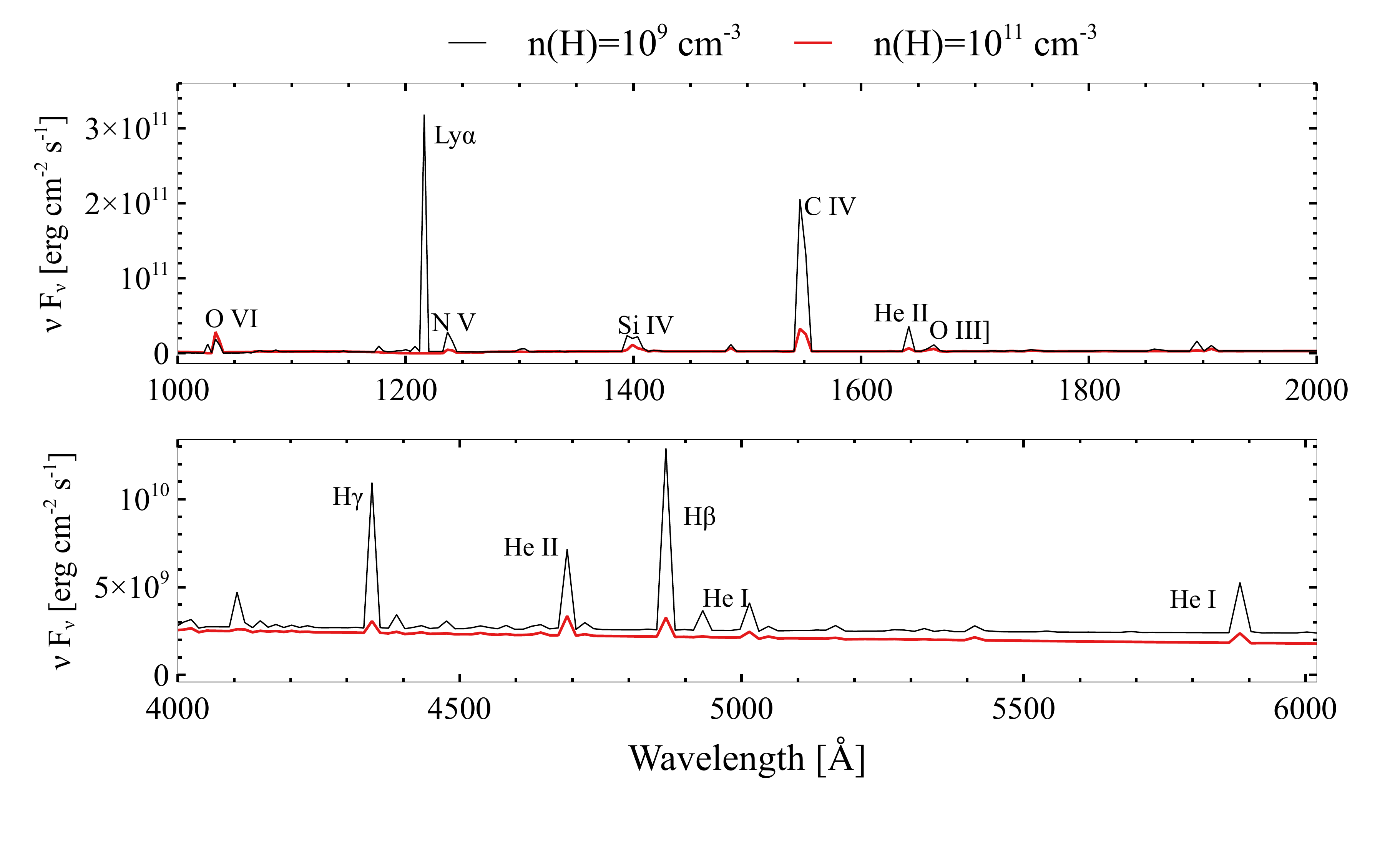}
 \caption{The total spectrum, including transmitted and reflected emission from the BLR for two densities of the equatorial obscurer. The upper panel shows the the UV regions and the lower panel shows the optical wavelengths. \edit1{This Figure shows how phenomena similar to changing-look AGN, 
 in which a Seyfert I  turns into a Seyfert II, would occur without changes in the
 intrinsic luminosity of the AGN}}\label{f5}
\end{figure*}

\begin{itemize}
    \item Case 1: In this low density regime (approximately $n($H$) <6\times 10^{9}$ cm$^{-3}$), the equatorial obscurer is transparent and has little effect on the SED or BLR. This may be the usual geometry in most AGN and results in a standard response of lines to the changes in the continuum luminosity. For low densities, the intervening wind has little effect on the optical/UV BLR, however, it does emit in other spectral ranges. This emission will be the subject of our future work. Changes in the EWs of the BLR emission lines follow the variations of the continuum luminosity.
    
    \item Case 2: \edit2{In this case} the obscurer has a higher density ($6\times 10^{9}$ cm$^{-3}$ to $4\times 10^{10}$ cm$^{-3}$). As Figure~\ref{f4} shows, for this range of hydrogen density, the BLR EW decreases independently of the AGN luminosity and the holiday occurs. Large changes in EW at $n$(H) = $6\times 10^{9}$ cm$^{-3}$ are due to the He$^{2+}$ -He$^{+}$ ionization front reaching the outer edge of the \edit1{}wind.  Much of the SED \edit1{in the XUV region} is absorbed.  

Case 2 produces the emission-line holiday.  In this scenario, the obscurer's density increased only slightly above Case 1.  When the ionization front appears, there are significant changes in the transmitted SED and the BLR follows these changes. These changes are independent of the observed far-ultraviolet continuum longward of 912 \AA, so appear as a holiday.  

One check of this model of the holiday is the  $\sim 19\%$ deficit in \civ \  EW observed by G16.  A smaller deficit, $\sim 6\%$, was observed by \cite{Pei17} for H$\beta$.  Figure~\ref{f4} shows that only small changes in the density $(\sim 8\%)$ are needed to produce this \civ \  deficit. The change needed to produce the holiday is shown by the gray shaded area.  Our model predicts the largest deficits for \siIVoIV, \HeIIoIII,  and \civ \ EWs, with a smaller deficit for Ly$\alpha$ EW, and the smallest deficit for H$\beta$ EW. These predictions are in the same sense as the AGN STORM observations \citep[G16 \&][]{Pei17}.

\edit1{Mg II was not observed by the STORM campaign, however we report this line for future reference. 
The line is nearly constant when the obscurer is in Case 1 while in Case 2 it is slightly affected. 
This is reasonable, since we do not expect such a low-ionization line to be affected as much as \civ\ or other 
similar lines.}

\item Case 3: In this case, the obscurer has the highest density ($>4\times 10^{10}$ cm$^{-3}$) and most of the ionizing radiation is blocked.  
As Figure~\ref{f4} shows, many of the broad emission lines vanish. 
A dense equatorial obscurer provides a scenario to produce a ``changing-look'' quasar,
transitioning from Seyfert 1 to Seyfert 2. 
Figure~\ref{f5} compares the optical/UV BLR spectrum for Cases 1 and 3. 
The upper panel shows 
that UV broad lines are suppressed by the dense equatorial obscurer. 
The optical lines in the lower panel almost disappear. 
This Figure suggests
that dense disk winds could contribute to the changing-look AGN 
phenomenon, since changes in the equatorial obscurer can cause transitions between Seyfert 1 and 2 without affecting the optical / UV continuum. The LOS obscurer, if present, is transparent at those wavelengths.  
\edit2{This would remove BLR emission during times when the black hole
remained active, a different form of the changing-look phenomenon.}

\end{itemize}

\section{Discussion and summary}

    Various types of winds are commonly seen in AGN.  They launch from inner regions of the disk, so the geometry shown in Figure~\ref{f1} might be typical, but usually in the transparent state (Case 1). A nearly fully ionized wind does not have a dramatic effect on the SED or lines. 
   
    The observed holiday corresponds to a temporary change in the density of the wind. We suggests that \edit2{wind shielding} is usually happening, but for most of the time we just do not notice it, because the wind is transparent. Such shadowing can be the missing ingredient in many AGN models.  
   
   Our model requires that the normal state of the equatorial obscurer is one where the ionization front is near the outer radius of the \edit1{wind}.  The ionization-front location depends on \edit1{the wind's} parameters. This variation greatly affects the transmitted SED, as the wind density changes. The original \cite{Kaastra14} model of the LOS obscurer ($\log U\approx-2.8$ or $\log \xi=-1.2$ erg cm s$^{-1}$ ) has an H~ionization-front and strong absorption at the Lyman limit \citep{Arav15}. Later \cite{Cappi16} proposed $\log U\approx-1$ ($\log \xi=0.5-0.8$ erg cm s$^{-1}$ ) for the LOS obscurer, and our tests show that this obscurer transmits the Lyman continuum, corresponding to the blue line\edit2{, Case 2,} in Figure 3. This shows it is likely that the physical state of the equatorial obscurer is such that the H, He ionization fronts are near the outer edge of the \edit1{wind} so that small changes in the model affect its location. This is why small changes in the obscurer's density (Figure~\ref{f4}) can produce significant changes in the SED and result in the holiday.
   
   This model appears fine-tuned since it is sensitive to the location of the ionization front.   
   But this geometry has a physical motivation from dynamical stability arguments. 
   \cite{Math77} point out that radiatively driven clouds become Rayleigh-Taylor unstable near 
   ionization fronts so that the cloud tends to truncate at that point. 
   This happens because \edit1{the Lyman continuum} radiative acceleration depends on the ion density, so falls precipitously when the gas recombines. This instability provides a natural explanation for why the obscurer tends to have an ionization front near its outer edge.
  	
  	Although this paper discusses the emission-line holiday, a simultaneous holiday happened for higher-ionization narrow absorption lines \citep[D19]{Kriss19}. D19 show that changes in the CF of the LOS obscurer could be responsible for the absorption-line holiday. This obscurer is part of the same wind that produces the equatorial obscurer. 
  	\edit2{The density of the equatorial obscurer, the base of the wind, might
  	change  because of instabilities in the flow}. This produces the emission-line holiday, as shown in Figure~\ref{f4}. At the same time, it seems likely that injecting more mass from the base of the wind into our LOS causes the wind to produce a substantial flow and larger \edit1{wind}. This produces a larger CF for the LOS obscurer, producing the absorption-line holiday. So, a denser equatorial obscurer results in a more extensive LOS obscurer. In other words, the emission and absorption-line holidays are unified by the structure of the wind. This is the first physical model of the holidays observed in NGC 5548, and the relationship between them. 
    
  As Figures~\ref{f3} shows, the SED transmitted through Case~2 is stronger than Case 1 for \edit1{energies $\lesssim$ 1 eV. 
   In Case 3, the SED is stronger than Case~1 for energies $\lesssim$ 5 eV. 
    The emission is mainly due to hydrogen radiative recombination in the optical and NIR and \edit2{Bremsstrahlung} in the IR.}
   These show that a dense equatorial obscurer can be a source of \edit2{continuum}, even in Case~2.  Such emission could  explain \edit2{the 
   significant thermal diffuse continuum component spanning the entire UV–optical–near IR continuum''} discussed in \cite{Goad19} and may be the source of the non-disk optical continuum emission discussed 
  by \cite{Ferland90}, \cite{Shi95}, and \cite{Chel19}. The BLR itself is also a source of non-disk continuum emission \citep{Korista01}. 

To summarize, we have demonstrated, for the first time, a physical model by which several different phenomena are unified by the presence of the disk wind: an absorption-line holiday, an emission-line holiday,
non-disk emission from the inner regions, and a contributor to the changing-look phenomenon. This shows the importance of \edit2{“wind shielding”} , in which a wind partially blocks the continuum ionizing other clouds. 
Large CF required by previous models (e.g \citet{Korista00, Kaspi99, Goad93}) supports the idea the \edit2{“wind shielding”} is likely. 
It may be the missing ingredient in understanding many AGN phenomena.

\edit1{We came to a model in which an intervening obscurer filters the continuum striking emission and
absorption line cloud after consideration of how they respond to changes during the STORM campaign.
Many papers have considered cloud shadowing as an appropriate explanation for very different observations.
\cite{Mur95}'s study of  accretion disk winds from AGN found that a dense gas could block 
the soft X-ray and transmit UV photons. Shielding permits wind acceleration to high velocities.  
This wind produces smooth line profiles and has a covering fraction of $10\%$. 
\cite{Lei04}, suggested a  wind model in which the continuum filtered through the wind
would better fit her models of BLR emission. 
Finally, \cite{Shem15} reproduced the Baldwin effect by use of such filtering. 
As the STORM campaign demonstrated, and these previous investigation suggested, cloud shadowing is a key ingredient in the physics of inner regions of AGN and must be considered in future studies.}

\acknowledgments
\edit2{We thanks the anonymous reviewer for their very careful comments on our paper. }Support for {\it HST} program number GO-13330 was provided by NASA through a grant from the Space Telescope Science Institute, which is operated by the Association of Universities for Research in Astronomy, Inc., under NASA contract NAS5-26555.We thank NSF (1816537), NASA (ATP 17-0141), and STScI (HST-AR.13914, HST-AR-15018) for their support and Huffaker scholarship for funding the trip to Atlanta to attend the annual AGN STORM meeting, 2017. MC acknowledges support from NASA through STScI grant HST-AR-14556.001-A and STScI grant HST-AR-14286, \edit1{and also support from National Science Foundation through grant AST-1910687}. 
M.D.\ and G.F.\ and F. G.\  acknowledge support from the NSF (AST-1816537), NASA (ATP 17-0141),
and STScI (HST-AR-13914, HST-AR-15018), and the Huffaker Scholarship.
B.M.P. and G.D.R.\ are grateful for the support of the
National Science Foundation through grant AST-1008882 to
The Ohio State University. M.M. is supported by the Netherlands Organization for Scientific Research (NWO) through the Innovational Research Incentives Scheme Vidi grant 639.042.525.


\clearpage
\begin{thebibliography}
\references
\bibitem[Arav et al.(2015)]{Arav15}
Arav, N., Chamberlain, C., Kriss, G. A., et al. 2015, A\&A 577, 37 



\bibitem[Cappi et al.(2016)]{Cappi16}
Cappi, M., De Marco, B., Ponti, G., et al. 2016, A\&A 592, A27 


\bibitem[Chelouche et al. (2019)]{Chel19}
Chelouche, D., Pozo Nuñez, F., Kaspi, Sh. 2019, NatAs,3,251–257 







\bibitem[Dehghanian et al. (2019)]{Deh19}
Dehghanian, M., Ferland, G.J., Kriss, G.A., Peterson, B.M., et al. 2019, ApJ...877..119D


\bibitem[DeRosa et al.(2015)]{DeRosa15}
De Rosa, G., Peterson, B. M., Ely, J., et al. 2015, ApJ, 806:128




\bibitem[Edelson et al.(2015)]{Edelson15}
Edelson, R., Gelbord, J. M., Horne, K., et al. 2015, ApJ, 806, 129


\bibitem[Fausnaugh et al.(2016)]{Fausnaugh16}
Fausnaugh, M.  M., Denney, K. D., Barth, A. J., et al. 2016, ApJ, 821, 56 

\bibitem[Ferland et al. (1992)]{Ferland92}
Ferland, G. J., Peterson, B. M., Horne, K, et al. 1992,ApJ,387, 95

\bibitem[Ferland et al.(2017)]{Ferland17}
Ferland, G. J., Chatzikos, M., Guzm\'{a}n, F., et al. 2017, RMxAA, 53,385

\bibitem[Ferland et al. (1990)]{Ferland90}
Ferland, G. J., Korista, K. T., Peterson, B. M. 1990,ApJ,363L, 21F

\bibitem[Goad \&  Koratkar(1998)]{Goad98}
Goad, M. and Koratkar, A. 1998, ApJ, 495, 718G

\bibitem[Goad et al.(2016)]{Goad16}
Goad, M., Korista, K. T., De Rosa, G., et al. 2016, ApJ, 824, 11


\bibitem [Goad et al. (2019)]{Goad19}
Goad, M. R. ,Knigge, C.,Korista,K. T.,Cackett, E., et al. 2019, MNRAS, 10, 1093 

\bibitem[Goad et al.(1993)]{Goad93}
Goad, M., O'Brien, P. T., Gondhalekar,P.M.1993,MNRAS.263..149G

\bibitem[Kaastra et al.(2014)]{Kaastra14}
Kaastra, J., S., Kriss, G. A., Cappi, M., et al. 2014, Science, 345, 64

\bibitem[Kaspi\& Netzer (1999)]{Kaspi99}
Kaspi, Sh. Netzer, H. 1999, ApJ...524...71

\bibitem[Kriss et al.(2019)]{Kriss19}
Kriss, G.A., De Rosa, G., Ely, J., et al. 2019, ApJ in press.

\bibitem[Korista \& Goad (2000)]{Korista00}
Korista,K. T. \& Goad, M. R. 2000, ApJ, 536, 284 

\bibitem[Korista \& Goad (2001)]{Korista01}
Korista,K. T. \& Goad, M. R. 2001, ApJ, 553, 695 

\edit1{\bibitem[Leighly (2004)]{Lei04}
Leighly, K. M. 2004, ApJ, 611, 125}


\bibitem[Mathews \& Blumenthal (1977)]{Math77}
Mathews, W. G. \& Blumenthal, G. R.1977, ApJ, 214: 10


\bibitem[Mathur et al.(2017)]{Mathur17}
Mathur, S., Gupta, A., Page, K., et al. 2017, ApJ, 846:55

\bibitem[Mehdipour et al.(2015)]{Mehd15}
Mehdipour, M., Kaastra, J., S., Kriss, G. A., et al. 2015, A\&A, 575, 22

\bibitem[Mehdipour et al.(2016)]{Mehd16}
Mehdipour, M., Kaastra, J., S., Kriss, G. A., et al. 2016, A\&A, 588, 139 

\edit1{\bibitem[Murray et al.(1995)]{Mur95}
Murray, N., Chiang, J., Grossman, S. A., Voit, G, M. 1995, ApJ, 451, 498}

\bibitem[Osterbrock \& Ferland(2006)]{Osterbrock06}
Osterbrock D. E., \& Ferland G. J., 2006, 
{\em Astrophysics of Gaseous Nebulae and Active Galactic Nuclei}, 2nd ed., Univ. Science Books, CA, Herndon, VA

\bibitem[Pei et al.(2017)]{Pei17}
Pei, L., Fausnaugh, M.  M., Barth, A. J., et al. 2017, ApJ, 837: 131 












\edit1{\bibitem[Shemmer \& Lieber (2015)]{Shem15}
Shemmer, O., Lieber, S. 2015, ApJ, 805, 124}

\bibitem[Shields et al. (1995)]{Shi95}
Shields, J. C., Ferland, G. J., Peterson, B. M. 1995, ApJ, 441, 507

\bibitem[Starkey et al.(2017)]{Starkey17}
Starkey, D., Horne, K., Fausnaugh, M. M., et al. 2017, ApJ, 835: 65


\bibitem[Sun et al.(2018)]{sun18}
Sun, M., Xue, Y., Cai, Z.\& Guo, H. 2018, ApJ, 857:86



 \end{thebibliography}
 \end{document}